\documentclass[11pt]{article}
\usepackage{amsmath}
\usepackage{amssymb}
\usepackage{amsthm}
\usepackage{amsfonts}
\usepackage{comment}
\usepackage{color}
\usepackage{mathrsfs}
\usepackage{braket}
\usepackage{graphicx}
\usepackage[subrefformat=parens]{subcaption}
\usepackage{cite}
\usepackage{here}
\usepackage{tikz}
\usetikzlibrary{decorations.pathmorphing}
\usepackage{caption}

\usepackage[stable]{footmisc}

\makeatletter

\@addtoreset{equation}{section}
\makeatother

\begin{document}

\pagestyle{empty}

\title{\bf{Mass fluctuations in non-rotating BTZ black holes}}

\date{}
\maketitle
\begin{center}
{\large 
Hyewon Han\footnote{dwhw101@dgu.ac.kr}, Bogeun Gwak\footnote{rasenis@dgu.ac.kr}
} \\
\vspace*{0.5cm}

{\it 
Department of Physics, Dongguk University, Seoul 04620, Republic of Korea
}

\end{center}

\vspace*{1.0cm}
\begin{abstract}
{\noindent
We investigate the impact of oscillations of a black-hole mass around its average value on the three-dimensional black hole geometry. Drawing on a classical framework that conceptualizes fluctuations near an event horizon as mass variations, we introduce a model where the metric of a black hole, formed from the collapse of a massive null shell, exhibits oscillatory behavior in spherical modes. This dynamic is encapsulated by a non-rotating BTZ-Vaidya solution, characterized by the black hole mass fluctuating at a resonant frequency $\omega$ and a small amplitude parameter $\mu_0$. Using a perturbative approach, solutions to the null geodesic equation are determined up to the second order in $\mu_0$. The temporal fluctuations of the event horizon's location induce alterations in the thermodynamic variables' values. Upon calculating the time-averaged values, it is observed that the mean Hawking temperature experiences a slight decrease due to these fluctuations, while the mean entropy exhibits an increase, deviating from trends observed in four- and higher-dimensional spacetimes. Further, the study delves into the influence of these fluctuations on the trajectories of null rays near the horizon, ultimately reaching the anti-de Sitter boundary at late times. The analytical computation of the general solution for the perturbed rays up to the second order underscores the novel approach of this study in examining the effects of mass oscillations on black hole thermodynamics and geometry, contributing a unique perspective to the field.
}
\end{abstract}

\newpage
\baselineskip=18pt
\setcounter{page}{2}
\pagestyle{plain}
\baselineskip=18pt
\pagestyle{plain}
\setcounter{footnote}{0}

\section{Introduction}
Black holes are characterized by their intense gravitational fields as well as complex structures and stand at the forefront of gravitational theory research. According to general relativity, a black hole is defined as a region of spacetime encased by an event horizon, a null hypersurface from which neither matter nor light can escape. This concept was further elaborated through Hawking's demonstration that the area of an event horizon cannot diminish over time \cite{hawking1972black}, drawing a parallel with the second law of classical thermodynamics and suggesting that the surface area of a black hole is analogous to its entropy. The formulation of black hole entropy and the generalized second law of thermodynamics were pioneering contributions made independently by Bekenstein \cite{bekenstein2020black,bekenstein1973black,bekenstein1974generalized}. Moreover, the correlation between surface gravity and temperature was highlighted, establishing that surface gravity acts as an indicator of black hole temperature \cite{bardeen1973four}. Hawking further elucidated a mechanism by which the black holes could emit thermal radiation, proportional to their surface gravity, due to quantum fluctuations near the event horizon that facilitate the creation of virtual particle pairs, allowing for the observation of a positive energy flux in asymptotic regions \cite{hawking1974black,hawking1975particle}. These seminal findings laid the foundation for the field of black hole thermodynamics, attracting considerable attention toward unraveling the mysteries of quantum physics within strong gravitational fields.

York provided a dynamical perspective on the Hawking effect through a semi-classical approach \cite{york1983dynamical}, positing that black holes could undergo zero-point oscillations in their quasi-normal modes. He proposed a model wherein the metric fluctuations of a spherically symmetric black hole were driven by an infalling null fluid exhibiting small-amplitude and resonant-frequency fluctuations. This model accounted for particle tunneling associated with event horizon fluctuations and offered statistical insights into the entropy and thermal fluctuations of black holes. Employing this framework, Barrab$\grave{\mathrm{e}}$s et al. further explored the ramifications of these fluctuations on the geometric properties of black holes \cite{barrabes1999metric}, marking a significant advance in the dynamic understanding of black hole physics.

The fluctuations in the vicinity of an event horizon of a black hole are closely approximated by a Vaidya-type solution. Investigations have demonstrated that minor perturbations in the mass of the black hole lead to modifications of the surface, thermodynamic properties, and outgoing flux within the black hole spacetime. The effects of fluctuations induced by additional quantum fields were examined through the incorporation of stochastic variables \cite{barrabes2000stochastically}. This model of a fluctuating black hole has been extended to theories encompassing spacetime dimensions greater than four \cite{han2023metric} and those incorporating a negative cosmological constant \cite{Han:2023ckr}.

The anti-de Sitter (AdS) spacetime, characterized by a negative cosmological constant, constant negative curvature, and a spatial boundary, emerges as a solution to the Einstein field equations. Black holes within this spacetime are pivotal to the AdS/conformal field theory (CFT) correspondence, suggesting a correlation between the thermodynamics of $D$-dimensional AdS black hole spacetime and that of the $(D-1)$-dimensional gauge field on the AdS boundary \cite{maldacena1999large,gubser1998gauge,MR1633012,MR1646895,aharony2000large}. Consequently, the thermodynamic attributes of AdS black holes across four and higher dimensions have garnered extensive focus \cite{Cai:2013qga,Zou:2013owa,Wei:2014hba,Zhang:2015ova,Xu:2015rfa,Zou:2016sab,Gwak:2018akg,Hegde:2020xlv,Cong:2021fnf,Gwak:2021tcl,Bai:2022vmx}. The first discovery of an AdS black hole solution in three dimensions was made by Ba$\tilde{\mathrm{n}}$ados, Teitelboim, and Zanelli \cite{Banados:1992wn,Banados:1992gq}, noted for its well-defined physical parameters and properties analogous to those of four-dimensional black holes, both classically and quantum mechanically. The thermodynamics within this three-dimensional AdS black hole spacetime has been actively explored from various perspectives \cite{Ichinose:1994rg,Cai:1998ep,Govindarajan:2001ee,Vagenas:2001rm,Sarkar:2006tg,Akbar:2007zz,Modak:2008tg,Akbar:2011qw,Hendi:2015wxa,Dehghani:2017thu,Gwak:2020zht,Gwak:2020xwc,Fathi:2021eig,Huang:2021eby,Bai:2022uyz,Devi:2023fqa}. Additionally, insights into the states of its dual CFT have been derived from analyses on the quasi-normal modes of Bañados--Teitelboim--Zanelli (BTZ) black holes \cite{Cardoso:2001hn,Birmingham:2001hc,Birmingham:2001pj,Son:2002sd,Birmingham:2002ph,Birmingham:2003wa,Carlip:2005zn}.

In this work, we investigate the implications of fluctuations in three dimensions, motivated by the premise that gravitational effects vary with the dimensionality of spacetime. It is posited that metric fluctuations in lower dimensions may reveal distinct characteristics. The theory of gravity in three dimensions presents unique and intriguing phenomena, serving as a valuable framework for investigating fundamental features without the intricate mathematical complexity encountered in higher dimensions. We herein focus on the influence of mass fluctuations on the geometry of the non-rotating BTZ black hole, employing a three-dimensional model inspired by previous works \cite{york1983dynamical,barrabes1999metric}. Given the critical role of mass as a parameter in black hole spacetime, it is anticipated that fluctuations in mass will have explicit effects on the geometric properties. Employing a classical and straightforward approach, the null geodesic equation of the metric, oscillating around its mean value, is solved. This yields a solution that characterizes a perturbed event horizon and elucidates modified thermodynamic properties, which differ from those observed in higher-dimensional contexts. Additionally, a general solution for the perturbed rays diverging towards the AdS boundary is derived. Notably, while solutions in higher-dimensional spacetimes necessitate approximations or limits due to complex calculations, the BTZ spacetime permits the analytical derivation of a comprehensive solution without resorting to simplifications. To the best of our knowledge, these findings represent the first of their kind in a three-dimensional setting.

The remainder of this paper is organized as follows. Section 2 introduces the model of a fluctuating black hole within a three-dimensional AdS spacetime, alongside the derivation of null geodesic equations for perturbations. Section 3 presents a solution for a perturbed event horizon and details the calculation of thermodynamic variables for the fluctuating black hole. Section 4 explores the impact of fluctuations on the trajectories of propagating radial rays. The findings are summarized in Section 5. Throughout this study, the metric signature $(-,+,+)$ and its units are utilized, with $G=c=1$.

\section{BTZ black hole with a fluctuating mass}
We attempt to investigate the consequences of minor fluctuations within black hole geometries. Utilizing a fluctuating black hole model established by \cite{york1983dynamical}, previous studies have delved into the alterations in geometry and radiation properties induced by fluctuations across four and higher dimensions \cite{barrabes1999metric, han2023metric, Han:2023ckr}. Drawing upon a similar conceptual framework, the current study focuses on the dynamics of fluctuating geometries within a three-dimensional spacetime, distinguished by a negative cosmological constant $\Lambda$.

The gravitational action in $(2+1)$ dimensions is delineated as follows
        \begin{align} 
            S=\frac{1}{16 \pi} \int d^3 x \sqrt{-g} \left( R+ \frac{2}{L^2} \right) + S_{Boundary},
        \end{align}
where $g = \det (g_{\mu \nu})$ represents the determinant of the metric tensor, $R$ denotes the Ricci scalar, and $L=(-\Lambda)^{-1/2}$ signifies the curvature radius of the AdS spacetime. The investigation centers on a spherically symmetric black hole geometry, epitomized by the BTZ solution \cite{Banados:1992wn} with zero angular momentum. The metric of the global BTZ spacetime, characterized by $J=0$, is articulated in Schwarzschild coordinates as
        \begin{align} 
            ds^2=-\left(-M+\frac{r^2}{L^2}\right) dt^2 +\left(-M+\frac{r^2}{L^2}\right)^{-1} dr^2 +r^2 d \phi^2 ,
        \end{align}
where $M=8M_{BH}$ integrates the mass parameter associated with the black hole's total mass $M_{BH}$.

We delineate small fluctuations near a black hole's event horizon as variations in the mass, adopting a Vaidya-type solution for a classical analysis. Mirroring the four-dimensional framework \cite{york1983dynamical}, which utilized axisymmetric modes with a zero azimuthal index within a non-rotating background geometry, this research is confined to spherical oscillations. The metric for the Vaidya-BTZ black hole, expressed in ingoing Eddington--Finkelstein coordinates $(v,r)$, is derived from the following line elements \cite{Husain:1994xa,Ziogas:2015aja}
        \begin{align} \label{metric}
            ds^2=-\left(-m(v)+\frac{r^2}{L^2}\right) dv^2 +2dvdr +r^2 d \phi^2,
        \end{align}
where $v=t+r_*$ denotes the advanced time coordinate, and $r_*$ represents the radial tortoise coordinate. The time-dependent mass function $m(v)$ in the metric is defined as
        \begin{align} 
            m(v)=-1+\left[ M\left\{1+\mu_0 \sin(\omega v)\right\} +1\right]\vartheta(v),
        \end{align}
where $\mu_0$ represents a dimensionless amplitude parameter and $\omega$ denotes the resonant frequencies. The inclusion of the Heaviside step function $\vartheta(v)$ presumes the black hole's formation via the spherical collapse of a massive null shell at $v=0$. Therefore, the metric \eqref{metric} describes a vacuum AdS spacetime inside the shell for $v<0$, transitioning to a fluctuating AdS black hole spacetime post $v>0$. Substituting this metric into the Einstein field equations yields the following stress--energy tensor
        \begin{align} 
            T_{\mu \nu}=\frac{1}{16 \pi r} \left[\delta(v) \left\{M(1+\mu_0 \sin (\omega v))+1\right\}+\vartheta(v)\mu_0M\omega \cos(\omega v)\right] l_{\mu} l_{\nu},
        \end{align}
where $l_{\mu}=-\partial_{\mu}v$ identifies null vector fields tangent to incoming null geodesics. When $v>0$, the energy density of the incoming null fluid—serving as the metric's source—exhibits fluctuations around its mean value of zero.

The conformal diagram in Figure \ref{fig:diagram} illustrates the causal structure of the entire geometry under the condition of no fluctuations ($\mu_0=0$), delineating critical aspects of the spacetime configuration. The hypersurface $v=0$, denoted by a double line, signifies the collapse of a null shell, demarcating the transition from vacuum AdS spacetime to BTZ black hole spacetime. The diagram features a left vertical line, representing the regular origin of AdS spacetime within the shell, and a right vertical line, indicating the AdS boundary at infinity. In the domain where $v>0$, a zigzag line marks a singularity within the causal structure of BTZ spacetime, while a dashed line delineates an event horizon. The narrative follows a radial null ray that reaches the boundary at a significantly late retarded time $u$. Such a ray, propagating closely by the event horizon, exhibits high effective frequencies, enabling its backward temporal tracing under the geometric optics approximation. Illustrated by a thick solid line within the diagram, the ray's journey commences from the boundary at $v=V_0$, progressing towards the regular origin in the vacuum AdS spacetime inside the shell. Upon reflection at the origin, it diverges outward, crossing the null shell at $v=0$, and continues its path near the event horizon post $v>0$, eventually reaching AdS infinity at an exceedingly large $u$ value. The analysis abstains from discussing the ray's behavior post-reflection at the boundary and its subsequent approach towards the black hole's singularity.

Transitioning to a black hole geometry imbued with small fluctuations, these perturbations influence both the horizon and the trajectories of propagating rays. To equate the retarded time $u$, at which a ray reaches infinity in the fluctuating spacetime, to the corresponding $u$ value in a stationary spacetime devoid of fluctuations, the advanced time $v=V_0(u)$ marking the ray's departure from the boundary must be adjusted to a new value $v=V(u)$ within the fluctuating geometry. By establishing the relationship between $u$ and $v$, the investigation delves into the ramifications of these fluctuations on the spacetime fabric. 

\begin{center}
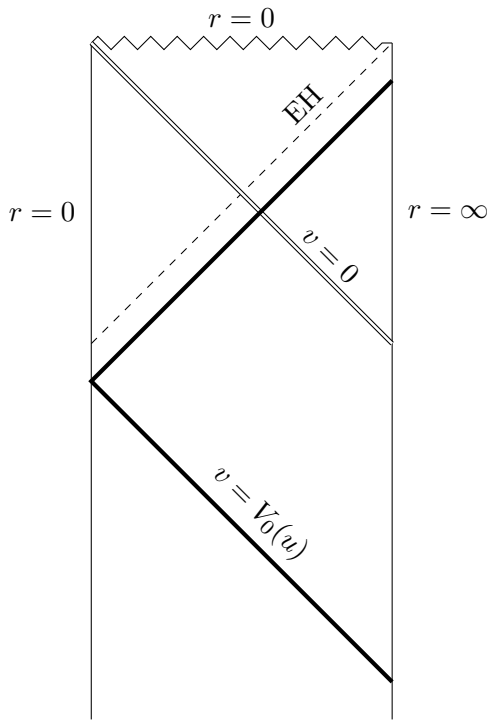

\begin{tikzpicture} 

    \node (I)    at (4,0)  {};
    \node (a)    at (4,-2) {};

    \path
     (I) +(90:4)  coordinate (Itop)
         +(-90:4) coordinate (Ibot)
         +(-90:5) coordinate (bot)
         +(180:4) coordinate (Ileft)
       ;
    \draw[dashed] (Ileft) -- node[near end, above, sloped] {EH}
                  (Itop);
    \draw (Itop) --      
          (bot) node[near start, right, inner sep=2mm]    {$r=\infty$} ;
      
    \path 
     (Itop) + (-4,0) coordinate (sing);
    \path 
      (sing) +(0,-9) coordinate (past);
    \draw (sing) -- (past) node[near start, left, inner sep=2mm]    {$r=0$};
    \draw[decorate,decoration=zigzag] (sing) -- (Itop)
         node[midway, above, inner sep=2mm] {$r=0$};

    \path
      (I) +(0,0) coordinate (v0);
    \draw[thin,double distance=1pt]
        (v0) -- 
          node[near start, above, sloped] {$v=0$}
        (sing);

    \path
      (Ileft) +(0,-0.5) coordinate (bounce);
    \path
     (Ibot)  +(0,-0.5) coordinate (Vu);
    \path
     (Itop) +(0,-0.5) coordinate (bounce2);
    \path
     (Itop) +(-0.5,0) coordinate (end);
    \draw[ultra thick]
     (Vu) -- node[midway, above, sloped] {$v=V_0(u)$} (bounce) -- (bounce2)  ;
    
\end{tikzpicture} 
\captionof{figure}{Conformal diagram depicting a non-rotating BTZ black hole resulting from the collapse of a null shell, in the absence of fluctuations.}
\label{fig:diagram}
\end{center}

To elucidate the impact of fluctuations on the event horizon and the propagation of rays within black hole geometry, it is imperative to solve the null geodesic equation for radially outgoing rays in the region $v>0$, 
        \begin{align} \label{ng}
            2\frac{dr}{dv}=-M\left[1+\mu(v)\right]+\frac{r^2}{L^2},
        \end{align}
where fluctuations are incorporated as $\mu(v)=\mu_0 \sin (\omega v)$. Given that $\mu_0$, the amplitude parameter, is minuscule for a massive black hole, a perturbation method is employed to facilitate the solution of this equation. The approach entails expressing the radial coordinate as follows
        \begin{align} \label{rc}
            r=r(v)=R(v)+\rho(v)+\sigma(v)+\cdots,
        \end{align}
where $R(v)$ denotes the radial coordinate in the absence of fluctuations, while $\rho(v)$ and $\sigma(v)$ represent the solutions to the first- and second-order perturbations in $\mu_0$, respectively. The analysis will limit consideration to second-order perturbations, disregarding terms of higher order.

To determine the unperturbed trajectory of an outgoing radial ray that reaches the AdS boundary of the black hole spacetime, the value of the retarded time $u=v-2R_*$ is specified, where
        \begin{align} \label{tortoise}
            R_*=\begin{cases}
                    \int \frac{1}{1+R^2/L^2} dR = L \arctan \left(\frac{R}{L}\right), & \mbox{for } v<0 \\
                    \int \frac{1}{-M+R^2/L^2} dR =\frac{L}{2 \sqrt{M}} \ln \left( \frac{R-\sqrt{M} L}{R+\sqrt{M} L} \right), & \mbox{for }v>0
                \end{cases}
        \end{align}
is the unperturbed tortoise coordinate. This setup ensures a clear framework for analyzing the path of radially outgoing rays.

The substitution of the radial coordinate expression \eqref{rc} into the null geodesic equation \eqref{ng} and its subsequent linearization yields 
\begin{align}
            2 &\, \frac{d R }{d v} = -M +\frac{R^2}{L^2}  , \label{zeroth} \\
            2 &\, \frac{d \rho }{d v} - \frac{2R}{L^2} \rho = - \mu M , \label{firstpe} \\ 
            2 &\, \frac{d \sigma }{d v} - \frac{2R}{L^2} \sigma = \frac{\rho^2}{L^2}. \label{secondpe}
        \end{align}
Solving these equations provides insights into the behavior of the event horizon within a fluctuating black hole spacetime. This methodology not only enables the determination of how thermodynamic properties are altered by fluctuations but also facilitates the derivation of a solution characterizing the trajectories of radial null rays as they propagate near the horizon and ultimately reach the AdS boundary. By examining these solutions, this study sheds light on the subtle effects that small perturbations have on the dynamics and thermodynamic properties of black hole spacetimes.

\section{Event horizon and thermodynamics in perturbed geometry}
In this section, the objective is to delineate a solution that accurately describes the position $r_H(v)=R_H(v)+\rho_H(v)+\sigma_H(v)+\cdots$ of an event horizon within a black hole spacetime subject to fluctuations. Initially, from the null geodesic equation, a null surface is identified where $r=R_H$ in the absence of fluctuations, representing the unperturbed horizon $R_H$ which satisfies the zeroth-order equation
        \begin{align}
            R_H^2=ML^2.
        \end{align}
The unperturbed surface gravity is defined as
        \begin{align}
            \kappa = \frac{R_H}{L^2}=\frac{\sqrt{M}}{L}.
        \end{align}
The equations governing the perturbations, \eqref{firstpe} and \eqref{secondpe}, utilize these variables to outline the perturbed position of the horizon. 
        \begin{align}
            \frac{d \rho_H}{d v} - \kappa \rho_H =& - \frac{\mu M }{2} , \\
            \frac{d \sigma_H}{d v} - \kappa \sigma_H =& \frac{\rho_H^2}{2L^2}.
        \end{align}
The solutions are found as
        \begin{align}
            \rho_H(v) = & \frac{\mu_0 M}{2 \kappa} \frac{ \Omega \cos (\omega v) + \sin (\omega v)}{1+ \Omega^2} ,\\
            \sigma_H(v) =& \frac{\mu_0^2 M}{16 \kappa} \frac{ (1-5\Omega^2) \cos (2\omega v) - 2\Omega (2-\Omega^2) \sin (2 \omega v) -   (1+\Omega^2)(1+4\Omega^2) }{(1+ \Omega^2)^2 (1+4\Omega^2)},
        \end{align}
where $\Omega=\omega/\kappa$ denotes a dimensionless frequency. Here the integration constants are set to preclude exponential growth in the perturbation terms over time $v$. The position of the horizon undergoes periodic variations contingent on the fluctuation parameters, with the second-order solution encompassing a double frequency $2\omega$. It is observed that the impact of fluctuations is magnified by an increase in the mass parameter $M$ and the curvature radius $L$.

These modifications in the black hole radius instigate alterations in thermodynamic variables. To capture the aggregate effects of fluctuations, the mean values of variables defined on the event horizon are computed by averaging over time $v$. The mean values of the surface area and the surface gravity on the perturbed horizon are formulated as follows
        \begin{align} 
            \overline{\mathcal{A}} &\equiv 2 \pi \, \overline{ r_H}= 2\pi \sqrt{M} L \left[1-\frac{\mu_0^2}{16(1+\Omega^2)} \right] = \mathcal{A}\left[1-\frac{\mu_0^2}{16(1+\Omega^2)} \right] ,  \label{area} \\
            \overline{\kappa} &\equiv \frac{\overline{r_H}}{L^2} = \frac{\sqrt{M}}{L}\left[1-\frac{\mu_0^2}{16(1+\Omega^2)} \right]=\kappa \left[1-\frac{\mu_0^2}{16(1+\Omega^2)} \right] , \label{surgra}
        \end{align}
where $\mathcal{A}$ and $\kappa$ denote the unperturbed values. Then we obtain the mean value of the Hawking temperature from a relation
        \begin{align} 
            \overline{T}_H = \frac{\hbar \, \overline{\kappa}}{2 \pi k_B} = T_H\left[1-\frac{\mu_0^2}{16(1+\Omega^2)} \right].  \label{HaT}
        \end{align}
where $k_B$ denotes the Boltzmann constant, and $T_H$ is the Hawking temperature in the fluctuation-free scenario. These variables, \eqref{area}--\eqref{HaT}, in the presence of fluctuations ($\mu_0 \neq 0$), exhibit diminished values compared to their stationary counterparts. This finding contrasts with outcomes in higher-dimensional spacetimes \cite{Han:2023ckr}, where fluctuations slightly increase the values of these variables. Additionally, the variations in the horizon area $\delta \mathcal{A} = \overline{\mathcal{A}} - \mathcal{A}$ and the Hawking temperature $\delta T_H = \overline{T}_H - T_H$ maintain the following relationship in three-dimensional spacetimes.
        \begin{align}
            \frac{\delta \mathcal{A}}{\mathcal{A}} = \frac{\delta T_H}{T_H},
        \end{align}
By identifying the energy of the system with the mean mass value of the black hole spacetime, the mean value of the Hawking--Bekenstein entropy can be derived from the first law of thermodynamics, $dE=\overline{T}_H d\overline{S}$, offering insights into the thermodynamic equilibrium in fluctuating black hole geometries.
        \begin{align} \label{entro}
            \overline{S} \simeq & \frac{k_B \overline{\mathcal{A}}}{4 \hbar} \left[1+\frac{\mu_0^2}{8(1+\Omega^2)} \right].
        \end{align}
The analysis reveals that the average entropy in spacetime with fluctuations is marginally increased, with this augmentation being directly proportional to the square of the amplitude parameter $\mu_0^2$. This observation presents a notable deviation from the behaviors identified in higher-dimensional spacetimes, where $D \ge 4$.

\section{Propagation of perturbed null ray}

Moving forward to solve the perturbation equations for the general case where $R > R_H$, the strategy involves utilizing the zeroth-order equation to transition from variable $v$ to $R(v;u)$, the unperturbed radial coordinate value determined by a given retarded time parameter $u$. This adjustment allows for the formulation of the perturbation equations in a unified manner
        \begin{align} \label{vc}
            \left(-M+\frac{R^2}{L^2} \right) \frac{d f}{d R}-\frac{2R}{L^2} f = 2F,
        \end{align}
where the first-order perturbation is given by
        \begin{align} 
            f=\rho, \qquad F=-\frac{\mu M}{2},
        \end{align}
and similarly, the second-order perturbation is described by
        \begin{align} 
            f=\sigma, \qquad F=\frac{\rho^2}{2L^2}.
        \end{align}
Solving these equations leads to the following expression
        \begin{align}
            f(R) = \left(-M+\frac{R^2}{L^2}\right) \left[-\int^{\infty}_{R}\frac{2 F(R')}{\left(-M+R'^2/L^2 \right)^2} \, d R' + f_{c} \right],
        \end{align}
where 
        \begin{align}
            f_{c} = \lim_{R \to \infty} \frac{f(R)}{\left(-M+\frac{R^2}{L^2}\right)}
        \end{align}
is an integration constant set to zero to ensure that the advanced time values for rays in the perturbed scenario match those in the unperturbed case. For convenience, dimensionless variables are defined as follows
        \begin{align}
            x=\frac{R-R_H}{R+R_H}, \qquad \tilde{u}=\kappa u, \qquad \tilde{f}=\frac{1}{R_H} f.
        \end{align}
In the case of $R>R_H$, $x>0$. Utilizing these variables, the solution for the first-order perturbation can be refined to
        \begin{align} 
            \tilde{\rho}(x)=\mu_0 \frac{x}{2(1-x)^2} I(x), \label{ff} 
        \end{align}
where
        \begin{align} 
            I(x)&=\int^1_x \frac{(1-\xi)^2}{\xi^2} \sin \left\{\Omega(\tilde{u}+\ln\xi)\right\} d\xi \nonumber \\
            &=\frac{2x\cos(\Omega \tilde{u})-\left\{2x-(x-1)^2 \Omega^2 \right\} \cos\left\{\Omega(\tilde{u}+\ln x)\right\}-(x^2-1) \Omega \sin \left\{\Omega(\tilde{u}+\ln x)\right\}}{x\Omega(1+\Omega^2)}.
        \end{align}
The solution for the second-order perturbation is articulated as
        \begin{align} 
            \tilde{\sigma}(x)=-\mu_0^2 \frac{x}{8(1-x)^2} \int^{1}_{x} \frac{I^2(\xi)}{(1-\xi)^2} d\xi. \label{ss}
        \end{align}
The integration of these expressions yields comprehensive results, as follows. 
        \begin{align} 
            \tilde{\sigma}(x)=&\frac{\mu_0^2}{16(1-x)^3\Omega^2(1+\Omega^2)^2(1+4\Omega^2)} \nonumber \\
            & \qquad \times \left[4x(1+x)(1+4\Omega^2) -(1+x)(1-x)^2\Omega^2(1+\Omega^2)(1+4\Omega^2) \right. \nonumber \\
            & \qquad \quad \left. +2x(1+x)(1+4\Omega^2) \left\{ \cos(2\Omega\tilde{u}) -2 \cos(\Omega \ln x)\right\} \right.  \nonumber \\
            &\qquad \quad\left. +\left\{x(1+x)(1+\Omega^2)(2+5\Omega^2)+(1+x^3)\Omega^2(1-5\Omega^2)\right\} \cos\left\{2\Omega(\tilde{u}+\ln x)\right\} \right. \nonumber \\
            & \qquad \quad\left. -4x(1+x)(1+4\Omega^2)\cos\left\{\Omega(2\tilde{u}+\ln x)\right\} -2x(1-x)\Omega^2(1+\Omega^2)(1+4\Omega^2) \ln x \right. \nonumber \\  
            & \qquad \quad\left. -x(1-x)\Omega(1-5\Omega^2)\sin(2\Omega\tilde{u}) + 4x(1-x)\Omega(1+4\Omega^2) \sin(\Omega \ln x) \right. \nonumber \\
            & \qquad \quad\left. -\left\{3x(1-x)\Omega(1+\Omega^2)(1+2\Omega^2)+2(1-x^3)\Omega^3(2-\Omega^2)\right\} \sin\left\{2\Omega(\tilde{u}+\ln x)\right\} \right. \nonumber \\
            &\qquad \quad \left. +4x(1-x)\Omega(1+4\Omega^2) \sin\left\{\Omega(2\tilde{u}+\ln x)\right\} \right]. 
        \end{align}
The full solution \eqref{rc} of the geodesic equation is obtained for the radially outgoing rays in the vicinity of the horizon within the fluctuating black hole spacetime. Remarkably, due to the intrinsic simplicity of three-dimensional gravity, it is feasible to deduce a complete solution for the perturbations through analytical computations, circumventing the need for approximations. This underscores the unique advantages of studying gravity in lower dimensions.

To elucidate the relationship between an advanced time $v$ when a radial ray departs from the boundary of AdS spacetime and a retarded time $u$ when it arrives at the boundary of a fluctuating black hole spacetime, we begin by examining the scenario devoid of fluctuations. In this context, a null ray traversing the region $v<0$ and intersecting the shell at $v = 0$ is governed by the following equation
        \begin{align} \label{v01}
            V_0=-2L \arctan \left(\frac{R_0}{L} \right),
        \end{align}
where $v=V_0$ represents the instance at which the ray departs infinity, and $R_0$ signifies the radial coordinate's value at $v=0$. Beyond $v = 0$, within the black hole geometry, the ray's movement adheres to a specific relation
        \begin{align} \label{u2}
            u=-\frac{L}{\sqrt{M}} \ln \left( \frac{R_0-\sqrt{M} L}{R_0+\sqrt{M}L} \right),
        \end{align}
where $u$ is the time at which it reaches the infinity. Given the continuity of the radial coordinate across the null shell, a functional relationship $V_0(u)$ emerges from integrating Equations \eqref{v01} and \eqref{u2}. With the introduction of fluctuations, while the rays continue to satisfy Equation \eqref{u2}, ensuring that both perturbed and unperturbed rays converge at identical $u$ values, the trajectory's portion influenced by fluctuations necessitates a modification of the equation for $V_0$; this results in
        \begin{align} \label{Vfunc}
            V=-2L \arctan \left(\frac{R_0+\rho(R_0)+\sigma(R_0)}{L} \right),
        \end{align}
where $v=V$ indicating the adjusted departure time $V$ for a perturbed ray. To deduce the function $V(u)$, solutions for the perturbations at the null shell are imperative. At $v=0$, the relationship becomes
        \begin{align} \label{shre}
            \tilde{u}+\ln x_0=0, 
        \end{align}
where $x_0$ denotes the $x$ value at the shell. Focusing on rays that reach the boundary at a considerably delayed time $\tilde{u}$, it is inferred that $x_0$ is minor for extensive $\tilde{u}$, hence justifying an approximation to first-order in $e^{-\tilde{u}}$. The solutions $\tilde{\rho}(x_0)$ and $\tilde{\sigma}(x_0)$ on the shell are derived and subsequently expressed in terms of $\tilde{u}$ using Equation \eqref{shre} as
        \begin{align} 
            \tilde{\rho}(\tilde{u})&=\frac{\mu_0}{\Omega(1+\Omega^2)} \left[ \frac{\Omega^2}{2}+e^{-\tilde{u}} \left\{ \cos (\Omega \tilde{u})-1 \right\} \right], \\
            \tilde{\sigma}(\tilde{u})&=- \frac{\mu_0^2}{\Omega^2(1+\Omega^2)^2} \left[ \frac{\Omega^4(5+2\Omega^2)}{8(1+4\Omega^2)}
            +e^{-\tilde{u}} \left\{ \frac{\Omega^4(5+2\Omega^2)}{4(1+4\Omega^2)} -\left(\frac{\cos (\Omega \tilde{u})-1}{2} \right)^2 \right. \right. \nonumber \\
            &\left. \left. \qquad\qquad\qquad\qquad\qquad\qquad\qquad +\frac{\Omega(1-5\Omega^2)}{16(1+4\Omega^2)}\sin(2\Omega\tilde{u})-\frac{\Omega^2(1+\Omega^2)}{8}\tilde{u} \right\} \right].
        \end{align}
Incorporating these outcomes into the Equation \eqref{Vfunc} yields
        \begin{align} 
            V(\tilde{u})&=-2L \arctan \left[\frac{R_H}{L}\left\{1+2 e^{-\tilde{u}}+\tilde{\rho}(\tilde{u})+\tilde{\sigma}(\tilde{u}) \right\}\right] \nonumber \\
            &=-2L \arctan \left[\frac{R_H}{L}\left\{1+\mu_0 \frac{\Omega}{2(1+\Omega^2)}-\mu_0^2\frac{\Omega^2(5+2\Omega^2)}{8(1+\Omega^2)^2(1+4\Omega^2)} \right.\right. \nonumber \\
            & \left. \left. \qquad \qquad \qquad +e^{-\tilde{u}}\left(2+\mu_0 \frac{\cos(\Omega\tilde{u})-1}{\Omega(1+\Omega^2)}-\mu_0^2 \frac{\Omega^2(5+2\Omega^2)}{4(1+\Omega^2)^2(1+4\Omega^2)} \right.\right.\right. \nonumber \\
            & \left. \left. \left. \qquad \qquad \qquad +\mu_0^2\left(\frac{\cos(\Omega\tilde{u})-1}{2\Omega(1+\Omega^2)}\right)^2-\mu_0^2\frac{(1-5\Omega^2) \sin(2\Omega\tilde{u})}{16\Omega(1+\Omega^2)^2(1+4\Omega^2)}+\mu_0^2 \frac{\tilde{u}}{8(1+\Omega^2)} \right) \right\}\right].
        \end{align}
This comprehensive analysis shows the discernible effects of small fluctuations on the propagation of rays within the three-dimensional AdS black hole spacetime, particularly pronounced in the regime of a late retarded time. This detailed exploration underscores the intricate interplay between spacetime dynamics and quantum fluctuations, offering profound insights into the behavior of gravitational fields under perturbed conditions.

\section{Conclusions}
We investigated the impact of minor fluctuations within a three-dimensional AdS black hole spacetime. Drawing inspiration from four-dimensional models that classically address fluctuations in asymptotically flat black holes \cite{york1983dynamical,barrabes1999metric}, this study introduced a model for a fluctuating AdS black hole in three dimensions. The investigation focused on a simplified scenario where the mass of a spherical black hole, resulting from the gravitational collapse of a massive null shell, undergoes oscillations in spherical modes with a minuscule amplitude. This scenario is aptly represented by a non-rotating BTZ-Vaidya solution, and according to the Einstein field equations, the energy density exhibits fluctuations around a mean value of zero.

Through the resolution of the outgoing null geodesic equation up to the second order in amplitude parameter, solutions were derived for elucidating the behavior of radial rays traversing along the event horizon and those directed towards the boundary. The findings indicate that minor perturbations induce periodic variations in the horizon's position, with the magnitude of these changes determined by the amplitudes and frequencies of fluctuations. Concurrently, the thermodynamic variables exhibit fluctuations, where it was observed that the average Hawking temperature marginally decreases relative to the spacetime devoid of fluctuations, whereas the average entropy increases. Notably, these characteristics are diametrically opposed to those observed in higher-dimensional spacetime, $D \ge 4$, across both asymptotically flat and AdS contexts, unveiling a distinct and intriguing property unique to three-dimensional environments.

Moreover, the inherent simplicity of three-dimensional spacetime facilitated the derivation of a comprehensive analytical solution that characterizes a radial ray's propagation near the horizon, incorporating up to second-order perturbations. Using this analytical solution, a relationship $V(\tilde{u})$ was established, delineating the advanced time at which the ray departs from the boundary against the retarded time at its arrival at the AdS black hole boundary. This relationship encapsulates the trajectory modifications induced by fluctuations, underscoring the nuanced effects of such perturbations within a three-dimensional AdS black hole spacetime.

\vspace{10pt} 

\noindent{\bf Acknowledgments}

\noindent This research was supported by Basic Science Research Program through the National Research Foundation of Korea (NRF) funded by the Ministry of Education (NRF-2022R1I1A2063176) and the Dongguk University Research Fund of 2023. BG appreciates APCTP for its hospitality during completion of this work.\\

\bibliographystyle{jhep}
\bibliography{ref_v2}

\end{document}